\documentstyle[12pt,epsfig]{article}
\begin{document}

\title{Remarks on some vacuum solutions of scalar-tensor cosmological models}
\author{V. B. Bezerra$^{1}$, C. Romero$^{1}$, G. Grebot$^{2}$, \\
M. E. X. Guimar\~aes$^{2}$, L. P. Colatto$^{3,4}$ \\
\mbox{\small{1 .Departamento de F\'{\i}sica, Universidade Federal da
Para\'{\i}ba, Jo\~ao Pessoa, Pb, Brazil}} \\
\mbox{\small{2. Departamento de Matem\'atica, Universidade de Bras\'{\i}lia, Bras\'{\i}lia,
Brazil}}%
\\
\mbox{\small{3. Coordena\c c\~ao de Part\'{\i}culas e Campos,}} \\
\mbox{\small{  Centro Brasileiro de 
Pesquisas F\'{\i}sicas, Rio de Janeiro, RJ, Brazil }} \\
\mbox{\small{4. Grupo de F\'{\i}sica Te\'orica Jos\'e Leite Lopes,
Petr\'opolis, RJ, Brazil}} \\
\mbox{\small{\bf valdir@fisica.ufpb.br, cromero@fisica.ufpb.br,}}\\
\mbox{\small{\bf guy@mat.unb.br , emilia@mat.unb.br , colatto@cbpf.br}}}
\maketitle
\date{}

\begin{abstract}
We present a class of exact vacuum solutions corresponding to de Sitter and
warm inflation models in the framework of scalar-tensor cosmologies. We show that 
in both cases the field equations reduce to planar dynamical systems with constraints.
Then, we carry out a qualitative analysis of the models by examining the phase diagrams
of the solutions near the equilibrium points.
\end{abstract}

\section{Introduction}

Scalar-tensor theories of gravity\cite{ST1,ST2} represent the most natural
alternatives to Einstein's theory of general relativity. The simplest and
earliest scalar-tensor theory\cite{ST1} considered a massless scalar field
and was formulated by using as basic metric tensor the physical tensor $%
\tilde{g}_{\mu \nu }$, to which matter is universally coupled (Jordan-Fierz
frame). Later, these theories were generalized\cite{ST2} by introducing a
scalar field self-interaction and a dynamical coupling to matter.

In what concerns cosmology, the presence of a scalar field - which from now
on we will call generically {\it dilaton}, has gravitational-strength
couplings to matter which violate the equivalence principle. To avoid
conflicts with experimental tests on the equivalence principle, it is
assumed that the dilaton acquires a mass large enough that any deviations
from Einstein's theory are confined on scales that are not sensitive on
cosmological scales. However, a lot of work\cite{kal} has been done in the
framework of low energy string theory - which is reminiscent of
scalar-tensor theories of gravity - focusing in the case where the scalar
field is massless. The mechanism which naturally reconciles a massless
dilaton with experimental test was proposed in Ref.\cite{DP}. Indeed, a
massless dilaton is shown to obey a Minimal Coupling Principle, e.g., to
decouple from matter by cosmological attraction in much the same way as the
generic attractor mechanism of the scalar-tensor theories of gravity\cite{DN}%
.

On the other hand, recent observational data which contain evidence for an
accelerated expansion of the universe\cite{riess} indicate that this may be
induced by scalar fields which appear naturally in scalar-tensor models.
Therefore, it is important to consider various possibilities of cosmological
scenarios in order to study, for example, the asymptotic behaviour at late
times of Friedmann-Robertson-Walker (FRW) cosmological models and to
investigate if the universe evolves towards a state indistinguishable or not
from the one predicted by general relativity \cite{santiago}.

The aim of this paper is to study vacuum solutions in the context of FRW
cosmologies with flat spatial curvature ($k=0$) for the cases of de Sitter
models and warm inflation\cite{freese}. It turns out that these two cases
lead to a class of field equations that can be written as planar dynamical
systems plus a constraint equation.

This work is outlined as follows. In section 2, we briefly describe the
scalar-tensor cosmology. In sections 3 and 4, we find the solutions of the
field equations for the de Sitter and warm inflation models, respectively.
Finally, in section 5, we present some final remarks concerning our results.

\section{Scalar-Tensor Cosmology: A Brief Review}

In this section we will make a brief review of the scalar-tensor
cosmological models and write out the field equations which we are going to
deal with in the next section. We start by considering the most general
scalar-tensor theories of gravity in the Jordan-Fierz frame which is given
by the action

\begin{equation}
{\cal S}= \frac{1}{16\pi} \int d^4 x \sqrt{-\tilde{g}} \left[\tilde{R} {\Phi}
- \frac{\omega({\Phi})}{\Phi}\partial^{\mu}{\ \Phi} \partial_{\mu}{\Phi} - 2{%
\tilde{\Lambda}}(\Phi)\right] + {\cal S}_{m}[\Psi_m , \tilde{g}_{\mu\nu}] ,
\label{JF}
\end{equation}
where $\tilde{g}_{\mu\nu}$ is the physical metric, $\tilde{R}$ is the
curvature scalar associated to it, ${\tilde{\Lambda}}(\Phi)$ is a
cosmological term which corresponds to the scalar field potential and ${\cal %
S}_{m}$ is the action for general matter fields which, at this point, is
left arbitrary.

The physical frame of Jordan-Fierz has the disadvantage of featuring
complicated evolution equations for the gravitational and scalar fields and
for this reason it is more convenient to work in the Einstein (conformal)
frame, in which the scalar and tensor degrees of freedom do not mix. Now,
let us define two new variables: $g_{\mu\nu}$, the metric tensor in the 
Einstein frame, and the scalar field $\phi$. Thus, making the transformation

\begin{equation}  \label{conf}
\tilde{g}_{\mu\nu} = A^2(\phi) g_{\mu\nu},
\end{equation}
we decouple the two modes of propagations. This relation tells us that the
metric tensor in the Einstein frame is conformally related to the physical
metric tensor in Jordan-Fierz frame. Besides, by a redefinition of the
following quantities 
\begin{eqnarray}
\Phi &=& \frac{1}{G A^2(\phi)},  \nonumber \\
\Lambda(\phi) & = & A^4(\phi){\tilde{\Lambda}}, \\
\alpha(\phi) & = & \frac{d \ln A(\phi)}{d\phi},  \nonumber
\end{eqnarray}
where $G$ is the bare gravitational constant and, by imposing the constraint,

\begin{equation}  \label{alp}
\alpha^2(\phi) = \frac{1}{[2\omega(\phi) + 3]},
\end{equation}
the new quantities $g_{\mu\nu} , \;\; \phi \;\; \mbox{and} \;\; A(\phi)$ are
uniquely defined in terms of the original quantities ${\tilde g}_{\mu\nu} ,
\Phi , \omega(\Phi)$. In the Einstein frame, the action (1) turns into 
\begin{equation}
{\cal S} = \frac{1}{16\pi G} \int d^4x \sqrt{-g} \left[ R - 2g^{\mu\nu}
\partial_{\mu}\phi \partial_{\nu}\phi - 2\Lambda(\phi)\right] + {\cal S}_m
[\Psi_m , A^2(\phi)g_{\mu\nu}],  \label{EF}
\end{equation}

In this new frame the field equations read

\begin{eqnarray}  \label{eqs}
G_{\mu\nu} + g_{\mu\nu}\Lambda(\phi) & = & 8\pi G T_{\mu\nu} + 2(\phi_{,
\mu} \phi_{, \nu} - \frac{1}{2} g_{\mu\nu}\phi^{, \sigma}\phi_{, \sigma} ), 
\nonumber \\
\Box_g \phi -\frac{1}{2} \frac{d\Lambda (\phi)}{d\phi} & = & - 4\pi G
\alpha(\phi) T,
\end{eqnarray}
where 
\[T_{\mu\nu} = \frac{2}{\sqrt{-g}}\frac{\delta {\cal S}_m}{\delta g^{\mu\nu}}, \]
with 
\[T^{\mu}_{\nu ; \mu} = \alpha(\phi) T \partial_{\nu} \phi , \]
which means that the energy-momentum tensor in the conformal frame is no
longer conserved, differently from the Jordan-Fierz frame in which the
energy-momentum tensor is conserved.

In what follows we will concentrate on the Friedmann-Robertson-Walker (FRW)
cosmologies with flat spatial curvature ($k=0$) and perfect-fluid matter
distributions. These models are represented by a spacetime with metric

\begin{equation}  \label{frw}
ds^2 = - dt^2 + R^2(t) [ dr^2 + r^2(d\theta^2 + \sin^2 \theta d\varphi^2)],
\end{equation}
being sourced by an energy-momentum tensor corresponding to a perfect fluid
given by

\begin{equation}
T^{\mu \nu }=(\rho +p)u^{\mu }u^{\nu }+pg^{\mu \nu },  \label{em}
\end{equation}
with $u^{\mu }\equiv \frac{dx^{\mu }}{d{\tau }}$ denoting the 4-velocity of
the fluid in the Einstein frame. We can relate quantities such as density
and pressure in both frames through the equations

\[
\rho = A^4(\phi) \tilde{\rho} \;\;\; \mbox{and} \;\;\; p = A^4(\phi) \tilde{p%
}. 
\]

For the FRW models, equations (\ref{eqs}) become 
\begin{eqnarray}  \label{eqs2}
- 3 \frac{\ddot{R}}{R} & = & 4\pi G (\rho + 3p) + 2(\dot{\phi})^2 -
\Lambda(\phi),  \nonumber \\
3 \left( \frac{\dot{R}}{R}\right)^2 & = & 8\pi G \rho + (\dot{\phi})^2 +
\Lambda(\phi), \\
\ddot{\phi} + 3\frac{\dot{R}}{R}\dot{\phi} & = & - 4\pi G \alpha(\phi) (\rho
- 3p) - \frac{1}{2} \frac{d\Lambda}{d\phi}.  \nonumber
\end{eqnarray}

In this work we are going to study general solutions of the above equations
for two particular vacuum inflationary models: the de Sitter model
(corresponding to $\Lambda(\phi) = \Lambda_0 = \mbox{constant}$) and the
warm inflation model \cite{freese} (corresponding to $\Lambda(\phi) = 3
\beta H^2$, with $H \equiv \dot{R}/R$). Let us define two new variables, $%
\psi$ and $H$, being given by 
\begin{eqnarray*}
\psi & \equiv & \dot{\phi} \\
H & \equiv & \frac{\dot{R}}{R}
\end{eqnarray*}

In terms of these two new variables, the equations (\ref{eqs2}) for the case
of de Sitter model can be written as 
\begin{equation}  \label{sit1}
\dot{H} = - H^2 -\frac{2}{3}\psi^2 + \frac{\Lambda_0}{3},
\end{equation}

\begin{equation}  \label{sit2}
\dot{\psi} = - 3H \psi,
\end{equation}

\begin{equation}  \label{sit3}
H^2 = \frac{\psi^2}{3} + \frac{\Lambda_0}{3},
\end{equation}
with the last equation being a constraint equation.

On the other hand, for the warm inflation model we have 
\begin{equation}  \label{warm1}
\dot{H} = (\beta - 1) H^2 -\frac{2}{3}\psi^2,
\end{equation}

\begin{equation}  \label{warm2}
\dot{\psi} = (2\,\beta- 3)H\,\psi-3\,\beta(\beta-1)\frac{H^3}{\psi},
\end{equation}
with the constraint equation given by

\begin{equation}  \label{warm3}
H^2 = \frac{\psi^2}{3(1-\beta)}.
\end{equation}
Thus, for the cases of de Sitter and warm inflation models, the field
equations (9) can be written as a planar dynamical system plus a constraint
equation in the variables $H$ and $\psi$, as we can conclude from the
analysis of Eqs.(\ref{sit1})-(\ref{sit3}) and (\ref{warm1})-(\ref{warm2}),
respectively.

\section{The de Sitter Model in Scalar-Tensor Cosmologies}

Let us now find analytical solutions of Eqs.(\ref{sit1})-(\ref{sit3}). From
Eq.(\ref{sit1}) and the constraint Eq.(\ref{sit3}) we get

\begin{equation}  \label{sit4}
\dot{H} = \Lambda_0 - 3 H^2,
\end{equation}

which can be integrated giving

\begin{equation}  \label{sol1}
H_1= \frac{\sqrt{\frac{\Lambda_0}{3}} \tanh(\sqrt{3 \Lambda_0}(t-t_0)) + H_0%
}{1 + \sqrt{\frac{3}{\Lambda_0}} H_{0}\tanh(\sqrt{3 \Lambda_0}(t-t_0))},
\end{equation}

\begin{equation}  \label{sol2}
H_2= \frac{\sqrt{\frac{-\Lambda_0}{3}} \tan(\sqrt{-3 \Lambda_0}(t-t_0)) + H_0%
}{1 - \sqrt{\frac{-3}{\Lambda_0}} H_{0}\tan(\sqrt{-3 \Lambda_0}(t-t_0))},
\end{equation}

\begin{equation}  \label{sol3}
H_3=\frac{H_0}{1 + 3H_{0}(t-t_0)},
\end{equation}
for $\Lambda > 0$, $\Lambda < 0$, $\Lambda = 0$, respectively, with $H_0$
being an integration constant. From Eqs.(\ref{sit1})-(\ref{sit3}) and (\ref
{sol1})-(\ref{sol3}) we obtain the solutions $\psi(t)$ for each value of $%
\Lambda_0$. Thus, we have the following solutions $(H_{i}(t),
\psi_{i}^{\pm}=\pm\sqrt{3H_{i}^{2} - \Lambda_0})$, for $i=1,2$, and $%
(H_{3}(t), \psi_{3}=\pm\sqrt{3}H_{3})$.

Let us note that the constraint equation (\ref{sit3}) is compatible with (%
\ref{sit1}) and (\ref{sit2}). The same remark applies to the equations (\ref
{warm1})-(\ref{warm3}). We conclude that the constraint equations (\ref{sit3}%
) and (\ref{warm3}) are nothing but particular curves of the set of integral
curves of the vector fields defined by the right-hand side of Eqs.(\ref{sit1}%
), (\ref{sit2}), (\ref{warm1}) and (\ref{warm2}). It can be directly
verified that the first integral of the dynamical system formed by (\ref
{sit1}) and (\ref{sit2}) is given by

\begin{equation}  \label{eq1}
H^2 = a\psi^{2/3} + {\frac{1}{3}}(\psi^2 + \Lambda_0)
\end{equation}
with $a=0$ corresponding to the constraint equation (\ref{sit3}).

Now, if a dynamical system has critical points (equilibrium points), it is
often useful to investigate the behaviour of the solutions near these
points. For $\Lambda >0$, Eqs.(\ref{sit1})-(\ref{sit3}) admit four critical
points in the phase plane $H\psi $. These points are located at $A(\sqrt{%
\frac{\Lambda _{0}}{3}}),0)$; $B(-\sqrt{\frac{\Lambda _{0}}{3}},0)$; $C(0,%
\sqrt{\frac{\Lambda _{0}}{2}})$; $D(0,-\sqrt{\frac{\Lambda _{0}}{2}},0)$.
These four points themselves represent solutions of the dynamical system
formed by Eqs.(\ref{sit1}) and (\ref{sit2}), however only $A$ and $B$
satisfy the constraint equation given by (\ref{sit3}). Incidentally, note
that the solutions represented by $A$ and $B$ are obtained from Eq.(\ref
{sol1}) by assigning the values $H_{0}=\pm \sqrt{\frac{\Lambda _{0}}{3}}$.
Moreover, $A$ and $B$ correspond(in the Einstein frame) to de Sitter
cosmological models whose scale factors are given, respectively, by

\begin{equation}  \label{eq2}
R(t)=R_{0}\exp({\sqrt{\frac{\Lambda_0}{3}} t}),
\end{equation}

\begin{equation}
R(t)=R_{0}\exp ({-\sqrt{\frac{\Lambda _{0}}{3}}t}),  \label{eq3}
\end{equation}
where $R_{0}$ is a constant. For $\Lambda <0$ the system has no critical
points, while for $\Lambda =0$ there is only one critical point at the
origin O(0,0) of the phase plane, corresponding, modulo a rescaling of the
coordinates, to the Minkowski spacetime. In all the above configurations
represented by the critical points the scalar field $\phi $ is constant,
since $\psi =0$.

We now will draw the phase diagrams corresponding to the solutions given by
Eqs.(\ref{sol1})-(\ref{sol3}) plus those represented by the critical points.
The curves appearing in these diagrams represent parametric solutions $%
(H(t),\psi (t))$ evolving in time, in the Einstein frame. Of particular
interest are the constant solutions corresponding to the equilibrium points
since all the solutions considered are attracted or repelled by them. The
position of the equilibrium points, when they exist, depends on the values
assigned to $\Lambda _{0}$.

\begin{figure}[h]
\hspace{0.3cm}
\includegraphics{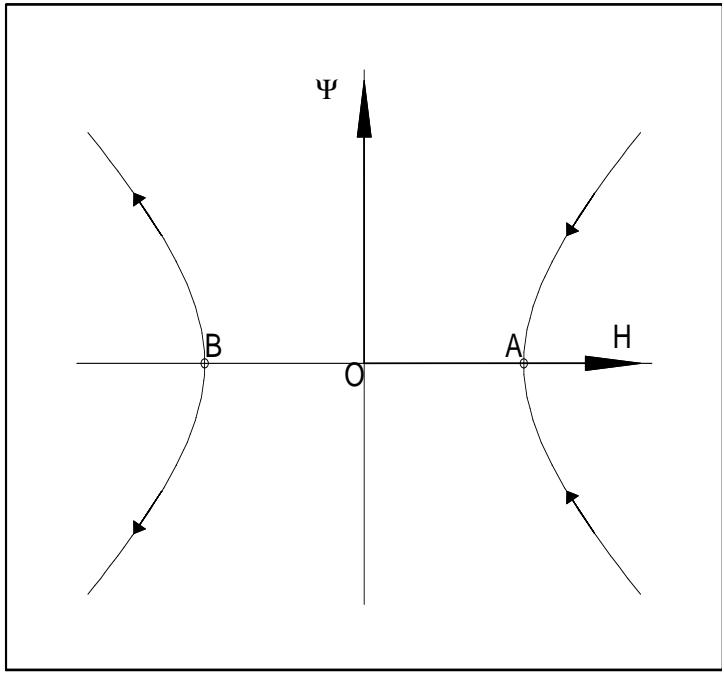}
\caption{ Phase diagram for $\Lambda _{0}>0.$ The critical points $A$
and $B$ represent de Sitter universes.}
\label{Fig1}
\end{figure}

The first diagram(see Fig. 1) refers to the case $\Lambda _{0}>0$. In this
case we have six solutions represented by the two critical points $(A$ and $B
$, the two curves lying in the upper part of the hyperbolae and the two
curves lying in the lower part. As we have seen before, the points $A$ and $B
$ describe expanding and collapsing universes in the Einstein frame (see
Eqs.(\ref{eq2}) and (\ref{eq3})). At this point, it is worth noting that since 
$\phi =constant$ in all solutions represented by the points $A$, $B$ and $O$%
, we see from (\ref{conf}) that the metric tensor $\tilde{g}_{\mu \nu }$ is
obtained from $g_{\mu \nu }$ by simply rescaling the coordinates. Thus, we
have essentially the same geometry in both Einstein and Jordan-Fierz frames.
It is also easy to see that, when $H_{0}>\sqrt{\frac{3}{\Lambda _{0}}}$
the point $A$ acts as an attractor for the two solutions $(H_{1}(t),\psi
_{1}^{+})t))$ and $(H_{1}(t),\psi _{1}^{-})t))$ when $t\rightarrow \infty $.
By reasons of continuity the same qualitative behaviour is carried over into
the Jordan-Fierz frame. Quite analogously, when $H_{0}<-\sqrt{\frac{3}{%
\Lambda _{0}}}$, then both $(H_{1}(t),\psi _{1}^{+}(t))$ and $(H_{1}(t),\psi
_{1}^{-}(t))$ move away from $B$ as time goes by. For $\Lambda =0$ ( Fig. 2
) we have three solutions, which describe an expanding universe, a
collapsing universe and Minkowski spacetime (represented by the equilibrium
point at the origin). When $\Lambda <0$ ( Fig.3 ) we have no equilibrium
points and in the Einstein frame the solutions appear as bouncing universes
possessing an expansion stage followed by a collapsing era.

\begin{figure}[h]
\includegraphics{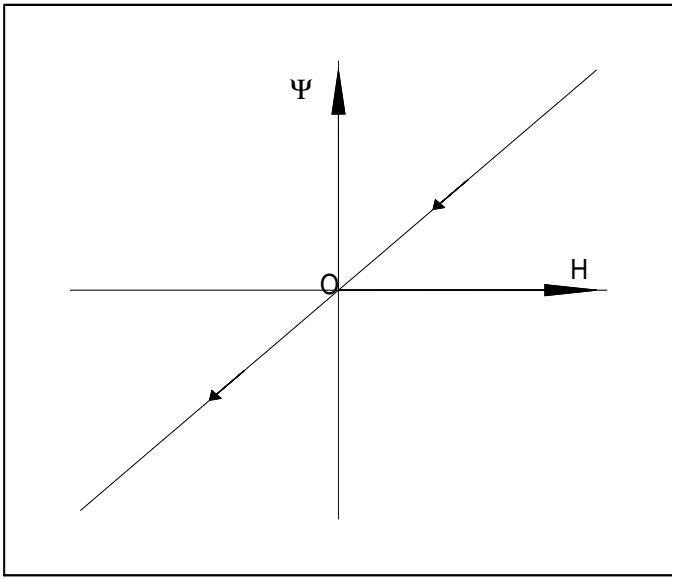}
\caption{ Phase
diagram for $\Lambda _{0}=0$. Here the critical points $A$%
and $B$ representing de Sitter universe
merge into the origin (Minkowski spacetime)} 
\label{Fig2}
\end{figure}

\begin{figure}[h]
\includegraphics{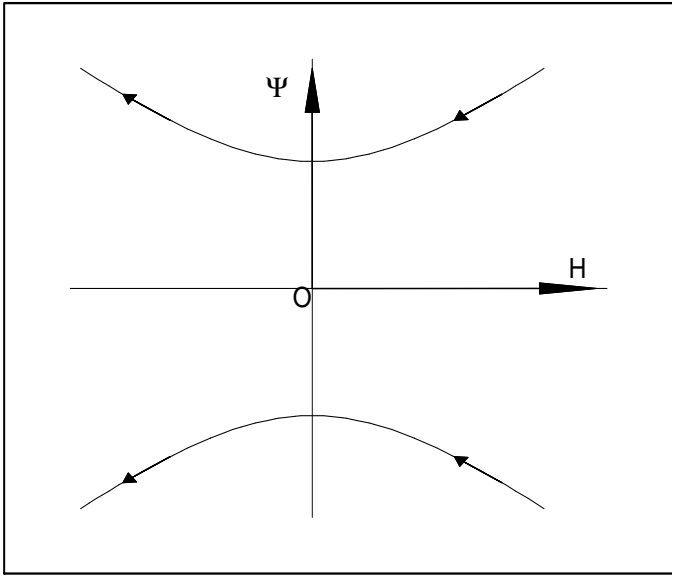}
\caption{ Phase diagram for 
$\Lambda _{0}<0.$ The critical points disappear.} 
\label{Fig3}
\end{figure}

\section{The Warm Inflation Model in Scalar-Tensor Cosmologies}

Let us now consider the case of warm inflation, which corresponds to the
system of equations (\ref{warm1})-(\ref{warm3}). Here, the solutions for $%
H(t)$ and $\psi(t)$ are easily obtained if we substitute Eq.(\ref{warm3})
into (\ref{warm1}). This leads to

\begin{equation}  \label{warm4}
\dot H= 3(\beta - 1)H^2
\end{equation}
the solution of which is given by

\begin{equation}  \label{warm5}
H(t)=\frac{H_0}{1 + 3H_{0}(1 - \beta)(t - t_0)},
\end{equation}
where $H_0$ is an integration constant. From the constraint equation (\ref
{warm3}) we have

\begin{equation}  \label{warm6}
\psi(t)= \pm {\sqrt{3( 1 - \beta)} H(t)}.
\end{equation}

Clearly the constraint equation (\ref{warm3}) also implies that $\beta <1$.
( The case $\beta =1$ leads to $\psi =0$ and $H=H_{0}$, which describes a de
Sitter universe).

\begin{figure}[h]
\includegraphics{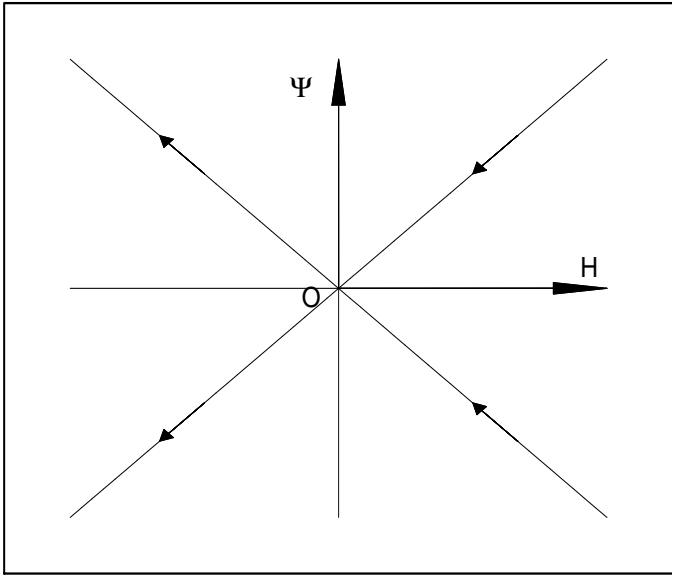}
\caption{Phase diagram for the warm inflation model.} 
\label{Fig4}
\end{figure}

The phase diagram corresponding to the solutions (\ref{warm5}) and (\ref
{warm6}) is shown in Fig.4. These solutions represent expanding and collapsing 
universes, for $H>0$ and $H<0$, respectively, and Minkowski spacetime ($H=\psi=0$). 
The expanding universe starts with a big-bang at $t^{*}=t_{0} + \frac{1}{3H_{0}(\beta - 1)}$ 
and approach Minkowski spacetime as $t \rightarrow \infty$, gradually slowing their
expansion rate. On the other hand, the collapsing models start at $t=-\infty$
as Minkowski spacetime and collapse at $t=t^{*}$. As we have noted in the
previous section, the same qualitative analysis of the solutions in the
vicinity of the origin may again be carried over into the Jordan-Fierz frame.

\section{Final remarks}

In this paper we have presented a class of exact solutions corresponding to
vacuum solutions of de Sitter and warm inflation models in the context of
scalar-tensor cosmology. These two particular models provide a class of
field equations that can be written as planar dynamical systems plus a
constraint equation.

In the case of the de Sitter model the dynamical system phase diagrams show
that if $\Lambda_{0} > 0$ and $\Lambda=0$ there exist solutions
corresponding to critical points (de Sitter universes). It just so happens
that in these cases the scalar field is constant. This fact allows us to
carry over our qualitative analysis of the solutions near the equilibrium
points from the Einstein frame to the Jordan-Fierz physical frame. The same
remarks applies to the warm inflation model, where we have only one
equilibrium point, which corresponds to Minkowski space-time. Finally, we
would like to stress that we do not touch here the problem of providing a
mechanism to terminate the inflationary phase of the universe, our solutions
for both models being valid for all $t$.

\section*{Acknowledgments}

V.B.B. and C.R. would like to thanks CNPq for partial financial support.
V.B.B., M.E.X.G. and L.P.C. would like to thank CAPES in the context of the
interinstitutional program PROCAD/CAPES for partial financial support.
M.E.X.G. would like to thanks the kind hospitality of the Departamento de
F\'{\i}sica of the Universidade Federal da Para\'{\i}ba where part of this work
has been developed.

\end{document}